\documentclass[11pt]{article}
\usepackage{epsfig}
\usepackage{latexsym}

\def\bg#1{\mbox{\boldmath$#1$}}

\newcommand{\sumint}{\hbox{$\sum$}\!\!\!\!\!\!\int}
\newcommand{\Tr}{\mbox{Tr}}

\newcommand{\del}{\partial}
\newcommand{\beq}{\begin{eqnarray}}
\newcommand{\eeq}{\end{eqnarray}}
\newcommand{\be}{\begin{eqnarray*}}
\newcommand{\ee}{\end{eqnarray*}}
\newcommand{\bk}{{\bf k}}
\newcommand{\bp}{{\bf p}}

\newcommand{\bx}{{\bf x}}

\newcommand{\ra}{\rightarrow}
\newcommand{\e}{\epsilon}

\newcommand{\nn}{\nonumber}

\newcommand{\ex}[1]{\langle\,#1\,\rangle}
\newcommand{\D}{{\cal D}}

\begin{document}

\centerline{\Large\bf {Quantum Corrections to the QED Vacuum Energy}}
\vskip 10mm
\centerline{Xinwei Kong and Finn Ravndal\footnote{Present address: NORDITA, Blegdamsvej 17,  
            DK-2100 Copenhagen \O, Denmark }} \bigskip
\centerline{\it Institute of Physics}
\centerline{\it University of Oslo}
\centerline{\it N-0316 Oslo, Norway}

\bigskip
{\bf Abstract:} {\small At energies much less than the electron mass $m$ the effects of quantum
fluctuations in the vacuum due to virtual electron loops can be included by extending the Maxwell 
Lagrangian by additional non-renormalizable terms corresponding to the Uehling and 
Euler-Heisenberg interactions. This effective field theory is used to calculate the properties 
of the QED vacuum at temperatures $T \ll m$. By a redefinition of the electromagnetic field,
the Uehling term is shown not to contribute. The Stefan-Boltzmann energy density is thus found 
to be modified by a term proportional with $T^8/m^4$ in agreement with the semi-classical 
result of Barton. The speed of light in blackbody radiation is smaller than one. Similarly, 
the correction to the energy density of the vacuum between two metallic parallel plates diverges 
like $1/m^4z^8$ at a distance from one of the plates $z \ra 0$. While the integral of the 
regularized energy density is thus divergent, the regularized integral is finite and corresponds 
to a correction to the Casimir force which varies with the separation $L$ between the plates 
as $1/m^4L^8$. This result is in seemingly 
disagreement with a previous result for the radiative correction to the Casimir force which 
gives a correction varying like $1/mL^5$ in a calculation using full QED.}
  
\vspace{10mm}
The use of effective field theories is steadily increasing in the investigation
of quantum phenomena at high energies\cite{Weinberg}\cite{Leutwyler}\cite{Kaplan}\cite{Manohar}. 
Since these theories  are generally non-renormalizable, it has taken some time before one has 
fully understood how they are supposed
to be used in higher orders of perturbation theory. Their applications are only valid below a 
given energy scale. Loop integrations are usually divergent and bring in high-energy degrees of 
freedom which are not described by the effective theory. This problem is cured by regularization.
It is most directly done by introducing an explicit momentum cutoff. Its effect can later be 
removed by the introduction of higher dimensional counterterms to cancel the divergences and 
yield finite results. An alternative and much more compact method is to use dimensional 
regularization. It seems to be ideally suited for effective field theories.

At energies of the order of the electron mass $m$, the most important interactions are
described by QED. In principle, one can also use this theory at much lower energies to explain
phenomena in atomic physics or condensed matter physics. But that would be technically very
difficult and one relies instead on other, effective theories involving more appropriate degrees
of freedom. The effects of high energy physics are then coded into the values of 
the coupling constants of new interactions. A typical example is NRQED\cite{NRQED} which has 
been developed during the last few years\cite{KL}. It makes it possible to calculate with the highest 
accuracy radiative and relativistic corrections to the energy levels of the simplest atoms such
as positronium\cite{Labelle} and muonium\cite{Nio}.

In the QED vacuum there are no physical electrons or photons. Since the photons are massless 
they will dominate all virtual processes for energies much less than the electron mass.
These quantum processes can therefore be described by an effective theory for the 
electromagnetic field where the effects of virtual electron loops appear as new, 
non-renormalizable interactions. The dominant terms are the Uehling corrections to the photon
propagator\cite{Uehling} and the Euler-Heisenberg four-photon interaction\cite{EH}. This effective
theory is discussed in the next section. We will show that the Uehling interactions have no
effects in the QED vacuum and can be transformed away by a redefinition of the photon field.
This simplification is further discussed in Appendix A. 
 
When the temperature is non-zero and below the electron mass $m$, we can use the effective theory
to calculate the free energy of the photon field. This is done in Section~2 where first
the radiative correction to the free  energy of interacting photons is derived. Using
thermodynamics we then also have the lowest correction to the Stefan-Boltzmann law for black 
body radiation. It results
from the Euler-Heisenberg interaction and is found to be in agreement with a previous, 
semi-classical calculation by Barton\cite{Barton}. In full QED this correction is due to very 
complicated 3-loop diagrams. On the other hand, in the effective theory it is 
directly given by a product of two 1-loop diagrams which are straightforward to evaluate. This
clearly demonstrates how powerful effective field theories can be for calculational purposes. 
The energy-momentum tensor
for the interacting system is constructed and its expectation value is found to be in 
agreement with the thermodynamic derivation. A few technical aspects of this calculation can be 
found in Appendix B. Also the temperature-dependent
corrections to the electric permittivity and magnetic susceptibility of the vacuum are calculated 
from the electromagnetic polarization tensor. It gives a reduction of the speed of light in
blackbody radiation in agreement with a recent result by LaTorre, Pascual and 
Tarrach\cite{LaTorre}. 

Here we consider only photons at low temperatures. In the opposite limit, i.e. $T \gg m$, one can 
also use the methods of effective field theory to simplify the calculations of thermodynamic 
properties\cite{BN}. The characteristic scale is then the temperature and the electron mass can be
set equal to zero. For length scales $>1/ T$ it is then safe to neglect the modes with non-zero 
Matsubara index and the effective theory becomes 3-dimensional. The coefficient of the 
Stefan-Boltzmann energy gets then
radiative corrections and there will be additional terms in the free energy which now varies
logarithmically with the temperature\cite{Jens}.

The Casimir effect for two parallel plates is considered in Section 3. After having constructed
the electromagnetic multipoles in the Coulomb gauge, the free field is quantized. Quantitative 
results for the field 
fluctuations are then given by the different correlators. In Appendix B they are also calculated
from the propagator of the photon field confined between two parallel
plates as constructed by Bordag, Robaschik and Wieczorek\cite{Bordag} in the Lorentz gauge. 
The correlators are in general found to be 
strongly dependent on the position between the plates. However, the vacuum energy density in the 
free theory is constant and gives the well-known Casimir force\cite{Casimir}\cite{Russian}
\cite{Milonni}. It has recently\cite{exp}
been measured with much greater precision than previously\cite{Sparnay}. However, when we 
include the Euler-Heisenberg interaction in the effective Lagrangian, we find that the 
regularized energy density diverges near the plates. It thus results in an infinite Casimir 
energy when integrated over the volume between the plates. This physical meaningless result
can be avoided by first integrating and then regularizing the Casimir energy. 
We then obtain a finite result for the correction to the Casimir force which is found to vary 
with the distance $L$ between the plates as $1/m^4L^8$. This is not in agreement with a
corresponding calculation using full QED  by Bordag {\it et al}\cite{Bordag} who
obtained a result varying like $1/mL^5$. Later this was confirmed by Robaschik, Scharnhorst and 
Wieczorek\cite{Robaschik} using a similar approach. 
If the Uehling term had not have been transformed away, it
would on the other hand have given a correction going like $1/m^2L^6$ for dimensional reasons. 
In general one should obtain the same results from both the full and 
the effective theory as long as they are properly matched.

It was first noted by Deutsch and Candelas\cite{DC} that the operations of integration and 
regularization do not in general commute in this type of problems. In order to analyze this
conflict more easily, a simpler model of an interacting, massless scalar field in 1+1 
dimensions has been investigated\cite{KR}. The model corresponds closely to the above 
description of interacting photons between two plates but without the uninteresting 
degrees of freedom in the transverse directions. It is then easy to show that the lack of
commutativity between integration and regularization is due to the breakdown of different
regularization schemes when one gets sufficiently close to one of the walls. In this region
unknown, cutoff-dependent effects show up reflecting the microscopic properties of the physical
walls.

\section{Interacting photons at low energies}

QED describes the interactions between the electromagnetic field $A_\mu(x)$ and the
electron field $\psi(x)$ by the standard Lagrangian
\beq
   {\cal L}(A,\psi) = -{1\over 4}F_{\mu\nu}^2 + \bar{\psi}[\gamma^\mu(i\del_\mu - eA_\mu) - m]\psi
\eeq
where $F_{\mu\nu} = \del_\mu A_\nu - \del_\nu A_\mu$ is the field strength. At
energies much below the electron mass $m$ and with no matter present, the first term which is the
Maxwell Lagrangian,
will dominate and describe free photons in the quantized theory. The effects of the
electrons are only felt through the coupling of photons to the virtual electron loops in the
vacuum. These interactions are suppressed by inverse powers of the electron mass compared with 
the Maxwell term and are thus very small.

\begin{figure}[htb]
 \begin{center}
  \epsfig{figure=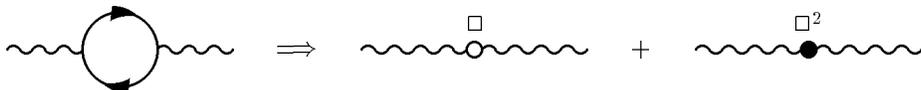}
 \end{center}
 \vspace{-8mm}            
 \caption{\small Uehling corrections to the photon propagator.}
 \label{fig1}
\end{figure}

\begin{figure}[htb]
 \begin{center}
  \epsfig{figure=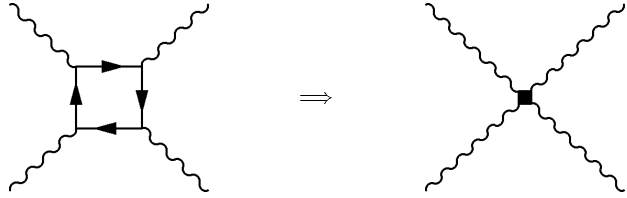,height=25mm}
 \end{center}
 \vspace{-8mm}            
 \caption{\small The Euler-Heisenberg four-photon vertex.}
 \label{fig2}
\end{figure}

One can include the effects of these virtual processes by extending the Maxwell Lagrangian to
include higher dimensional operators that describe these additional interactions. 
Lorentz invariance combined with parity and gauge invariance require these interactions to
be constructed from the electromagnetic field tensor $F_{\mu\nu}$ and derivatives thereof. The
coefficients of such operators in this effective Lagrangian ${\cal L}_{eff}$ can then be obtained
by matching results with those obtained from full QED\cite{NRQED}. In principle this
will lead to the same result as directly integrating out the electron field in the 
vacuum-to-vacuum transition amplitude or partition function,
\beq
     e^{i\!\int\!d^4x{\cal L}_{eff}(A)} = \int\!\D\psi\D\bar{\psi}\,
     e^{i\!\int\!d^4x{\cal L}(\psi,A)} 
\eeq
Since the QED Lagrangian is quadratic in the electron field, the integration gives a functional
determinant and the effective Maxwell Lagrangian can formally be written as
\beq
     {\cal L}_{eff}(A) = -{1\over 4}F_{\mu\nu}^2 - i\Tr\log{[\gamma^\mu(i\del_\mu - eA_\mu) - m]}
                                                    \label{trlog}
\eeq
It can expanded in an infinite series of operators ${\cal O}_n$ with increasing dimensions 
$D_n > 4$ preceeded by expansion coefficients proportional with $m^{4-D_n}$.
The lowest order contribution is due the to 1-loop correction of the photon propagator as shown in 
Fig.\ref{fig1}. It modifies the Coulomb potential at short distances. This is the Uehling 
interaction\cite{Uehling} and it is quadratic in the photon field. Expanding it to lowest order 
in the inverse electron mass it gives rise to the interaction\cite{IZ}
\beq
    {\cal L}_U = {\alpha\over 60\pi m^2}F_{\mu\nu}\Box F^{\mu\nu}                 \label{LU}
\eeq
where $\alpha = e^2/4\pi$ is the fine structure constant and $\Box \equiv \del_\mu\del^\mu$. 
Inserting extra photon lines into the electron loop gives ${\cal O}(\alpha)$ or higher corrections 
to the coefficients of these interactions which will be neglected in the following. Expanding
to higher orders in $1/m$, one finds in addition to the dimension $D=8$ Uehling term, the 
Euler-Heisenberg effective interaction\cite{EH}
\beq
     {\cal L}_{EH} = {\alpha^2\over 90m^4}\left[(F_{\mu\nu}F^{\mu\nu})^2 
                   + {7\over 4}(F_{\mu\nu}\tilde{F}^{\mu\nu})^2\right]           \label{LEH}
\eeq
where $\tilde{F}_{\mu\nu} = {1\over 2}\epsilon_{\mu\nu\rho\sigma}F^{\rho\sigma}$ is the dual 
field strength. It can be derived analytically from the functional determinant in (\ref{trlog})
when the field is constant as shown in many textbooks\cite{IZ}. This effective interaction gives a 
point coupling between four photons. In full QED it arises from the coupling of the photons to a 
virtual electron loop as shown in Fig.\ref{fig2}. 
To this order in the inverse electron mass we thus have the effective Lagrangian
\beq
     {\cal L}_{eff} = -{1\over 4}F_{\mu\nu}^2  + {\cal L}_U + {\cal L}_{EH}        \label{eff}
\eeq
for the interacting Maxwell field. The equation of motion for the free field is 
$\Box F_{\mu\nu} = 0$. It can now be used to simplify the interaction terms  which in 
our case with no matter means that the Uehling term ${\cal L}_U$ can be effectively set equal 
to zero. Since the equation of motion is only satisfied by on-shell photons, one may then 
question the validity of this simplification where the interactions are used to generate loop 
diagrams with virtual photons. A justification for this procedure is given in Appendix A where 
we for simplicity consider a similar
scalar theory which also couples to matter. Here it can be said that the effective Lagrangian 
is supposed to be used in the functional integral for the partition function 
where one is generally  allowed to shift integration variables by local redefinitions of the field. 
In fact,  we see that by the transformation
\beq
     A_\mu \ra A_\mu + {\alpha\over 30\pi m^2}\Box A_\mu                       \label{trans}
\eeq
the Uehling interaction is eliminated. The field redefinition does not modify the four-photon
(\ref{LEH}) interaction to this order in the electron mass.

The effective theory of photons at low energies based on the Euler-Heisenberg Lagrangian is
a non-renormalizable theory since the interaction has dimension $D  = 8$. Until recently it has
therefore been treated as a classical theory. It seems that Halter\cite{Halter} was the first to
consider it instead as a fullfledged quantum theory in which one consistently can calculate
radiative corrections. He considered in particular higher order corrections to photon-photon
scattering and calculated the contribution from one-loop  diagrams where the Euler-Heisenberg vertex 
appeared twice, i.e. giving an amplitude going like $\omega^8/m^8$ where $\omega$ is the photon 
energy. This corresponds to the single action of a dimension $D=12$ vertex and represents 
a subleading contribution\cite{Balholm}. More important are the ${\cal O}(\alpha)$ 
corrections to the tree-level result due to higher order corrections to the coefficient of the 
Euler-Heisenberg interaction (\ref{LEH}). These where originally calculated by Ritus\cite{Ritus}
and later confirmed by Reuter, Schmidt and Shubert\cite{RSS}. Secondly, as pointed out recently 
by Dicus, Kao and Repko\cite{DKR}, the dimension $D=10$ interaction in the expansion of the 
effective theory will become important at sufficiently high energies and thus also contribute 
at tree-level. This new interaction, which they have constructed, involves four photon field 
strengths with two derivatives. In the following we will ignore these higher order corrections.

\section{QED at low temperatures}
\begin{figure}[htb]
 \begin{center}
  \epsfig{figure=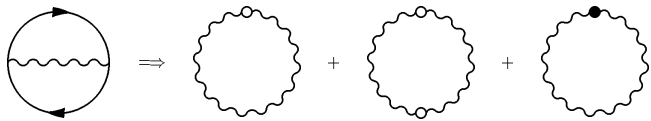, height=15mm}
 \end{center}
 \vspace{-8mm}            
 \caption{\small Possible Uehling correction to the free energy to order $1/m^4$.}
 \label{fig3}
\end{figure}

The energy density of free photons at non-zero temperature $T$ is given by the Stefan-Boltzmann law
\beq
      {\cal E} = {\pi^2\over 15}T^4                                               \label{SB}
\eeq
Due to interactions with virtual electron-positron pairs this fundamental formula will in general
have quantum corrections. The lowest contribution results from QED diagram in Fig.\ref{fig3} 
which can be obtained from the photon self energy in Fig.\ref{fig1} by connecting the two photon 
lines.
It is therefore the Uehling
contribution to the vacuum energy. However, when it is calculated for temperatures $T \ll m$ it is 
found to be exponentially suppressed like $\exp(-m/T)$\cite{Woloshyn}. This is consistent with our
observation that the Uehling interactions can be transformed away in the low-energy effective 
theory.

The first correction to the Stefan-Boltzmann law is therefore given by Euler-Heisenberg interaction 
and must for dimensional reasons vary with the temperature like $T^8/m^4$. We will calculate 
its absolute magnitude using the imaginary-time reformulation of the effective
theory. In addition to an over-all change of sign, the part involving the dual tensor in the 
effective Lagrangian will also change sign so that the we have the Euclidean theory 
\beq
   {\cal L}_{E} &=& {1\over 4}F_{\mu\nu}^2 - {\alpha^2\over 90m^4}\left[(F_{\mu\nu}F_{\mu\nu})^2 
                 - {7\over 4}(F_{\mu\nu}\tilde{F}_{\mu\nu})^2\right]      \label{LE1}  \\
                &=& {1\over 2}({\bf E}^2 + {\bf B}^2)  - {2\alpha^2\over 45m^4}
    \left[({\bf E}^2 + {\bf B}^2)^2  -  7({\bf E}\cdot{\bf B})^2\right]          \label{LE2}
\eeq
For the purpose of doing perturbation theory we need the correlator 
$\ex{F_{\mu\nu}(x)F_{\alpha\beta}(y)}$ of the free field. It follows from the free
photon propagator with 4-momentum $k_\mu = ({\bf k},k_4)$ in Euclidean space,
\beq
    \ex{A_\mu(k)A_\nu(-k)} = {1\over k^2}\left[\delta_{\mu\nu} 
                           + (\xi - 1){k_\mu k_\nu\over k^2}\right]                 \label{AA}
\eeq
where we can choose the gauge-fixing parameter $\xi = 1$. Then we have
\beq
    \ex{A_\mu(k)F_{\alpha\beta}(-k)} = {i\over k^2}
    \left[k_\alpha \delta_{\mu\beta} - k_\beta \delta_{\mu\alpha}\right]            \label{AF}
\eeq
which will be needed when we calculate the polarization tensor. The gauge invariant correlator of 
the field tensor is therefore
\beq
     \ex{F_{\mu\nu}(k)F_{\alpha\beta}(-k)} = {1\over k^2}
     \left[k_\mu k_\beta\delta_{\nu\alpha} - k_\mu k_\alpha\delta_{\nu\beta} 
     + k_\nu k_\alpha\delta_{\mu\beta} -  k_\nu k_\beta\delta_{\mu\alpha}\right]       \label{FF}
\eeq
When the photon field is in thermal equilibrium at temperature $T$, the fourth component of the
four-momentum vector $k_\mu$ is quantized, $k_4 = \omega_n = 2\pi T n$
where the Matsubara index $n$ takes all 
positive and negative integer values for bosons. Loop integrations are then done by the sum-integral
\beq
     \sumint_k = T \!\!\sum_{n=-\infty}^\infty \!\int{d^3 k\over (2\pi)^3}            \label{si}
\eeq
where the 3-dimensional momentum integration is dimensionally regularized and the Matsubara 
summation is regularized using zeta-functions. Many such sum-integrals have been calculated by 
Arnold and Zhai\cite{AZ} and Braaten and Nieto\cite{BN}. Some of the needed integrals are given in 
Appendix B. We notice here that integrals on the form $\sumint_k  k^{2n}$  are  non-zero only when 
$n < 0$. When $n \ge 0$ they look like being highly divergent, but are actually zero when using
dimensional regularization.

As a check, we can now obtain the energy density of free photons which in the Euclidean formulation
is given by
\beq
    {\cal E}(\bx)= -{1\over 2}\ex{{\bf E}^2(\bx) - {\bf B}^2(\bx)}                 \label{calE}
\eeq 
Using the basic integral 
\beq
    \sumint_k {k_\mu k_\nu \over k^2}  = {\pi^2 T^4\over 90}                           \label{sumint}
    (\delta_{\mu\nu} - 4\delta_{\mu 4}\,\delta_{\nu 4})
\eeq
we obtain from (\ref{FF}) above
\beq
    \ex{{\bf E}^2(\bx)} = \sumint_k {{\bf k}^2 + 3k_4^2\over {\bf k}^2 +k_4^2} 
                        = -  {\pi^2 T^4\over 15}                                   \label{EE}
\eeq
and
\beq
    \ex{{\bf B}^2(\bx)} = \sumint_k {2{\bf k}^2\over {\bf k}^2 +k_4^2} =  
                                                       {\pi^2 T^4\over 15}           \label{BB}
\eeq 
which reproduces the Stefan-Boltzmann energy (\ref{SB}). The  corresponding free energy density is 
\beq
       {\cal F} = - {\pi^2 \over 45}T^4           \label{Free}
\eeq
 and equals the negative of the pressure $P$.

\subsection{Corrections to the free energy}
\begin{figure}[htb]
 \begin{center}
  \epsfig{figure=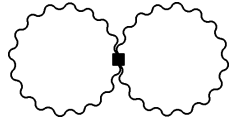,height=15mm}
 \end{center}
 \vspace{-8mm}            
 \caption{\small Euler-Heisenberg correction to the free energy.}
 \label{fig4}
\end{figure}

The free energy of interacting photons follows directly from the partition function 
\beq
    Z = e^{-\beta \int d^3x \;{\cal F}} 
      =  \int\!{\cal D}A_\mu \, e^{-\int_0^\beta\!d^4x\, {\cal L}_E}    \label{Z}
\eeq
where $\int_0^\beta\! d^4x \equiv \int_0^\beta dx_4 \int\!d^3x $. To lowest order in 
$T/m$ the correction to the free energy density is
thus simply given by the expectation value $\Delta{\cal F} = \ex{\Delta{\cal L}_E}$ in the free
theory where $\Delta{\cal L}_E$ is the  Euclidean Euler-Heisenberg interaction (\ref{LE1}). 
For the evaluation of the expectation value, it is convenient to rewrite it as
\beq
   \Delta {\cal F} = {\alpha^2\over 90m^4}\ex{7F_{\mu\nu}F_{\nu\beta}F_{\beta\alpha}F_{\alpha\mu}
   - {5\over 2}F_{\mu\nu}F_{\nu\mu}F_{\beta\alpha}F_{\alpha\beta}}
\eeq
after having expressed the dual field tensor in terms of the usual one.
We see that the correction to the free energy corresponds to the calculation
of the 2-loop Feynman diagram in Fig.\ref{fig4}. 
It is seen to be really the product of two 1-loop diagrams 
and therefore almost trivial to calculate compared with the corresponding 3-loop diagram in full 
QED. Using the correlator (\ref{FF}) and doing all the momentum contractions, we find the result 
to be given by the double sum-integral
\beq
    \Delta {\cal F} = - {22\alpha^2\over 45 m^4}\sumint_p \sumint_q \frac{(p\cdot q)^2}{p^2q^2}
\eeq
which can be evaluated from the basic integral (\ref{sumint}) which gives
\beq
     \sumint_p \sumint_q \frac{(p\cdot q)^2}{p^2q^2} = \frac{\pi^4T^8}{675}
\eeq
The free energy density of interacting photons to lowest order in perturbation theory is therefore
\beq
   {\cal F} = -{\pi^2\over 45}T^4 - {22\,\pi^4\alpha^2 \over 3^5\cdot 5^3}{T^8\over m^4}\label{F2}
\eeq
which is also the negative pressure.

This result has previously been derived by Barton\cite{Barton} using a semi-classical method and
treating the interacting photon gas as a material medium. Since the entropy is given by the 
derivative of the free energy with respect to the temperature, the energy density follows from 
${\cal E} = (1 - T\del/\del T){\cal F}$ as
\beq
    {\cal E} = {\pi^2\over 15}T^4 
             + {7\cdot 22\,\pi^4\alpha^2 \over 3^5\cdot 5^3} {T^8\over m^4}  \label{E}
\eeq
Obviously the corrections due to the Euler-Heisenberg interaction are totally negligible at 
ordinary temperatures. One can perhaps speculate that they may be of relevance under special 
astrophysical conditions. In this spirit Barton\cite{Barton} has investigated the implications 
of these photon interactions for the Planck distribution of blackbody radiation.

\subsection{The energy-momentum tensor}
The previous calculation was essentially based upon the thermodynamics of the interacting photon
gas. In particular we see that the factor seven which relates the correction to the energy to
the free energy comes from taking $1 - T\del/\del T$ acting on $T^8$ which seems to be of purely
thermodynamic origin. However, we also know that the energy density ${\cal E}$ and pressure $P$ are
both directly given by the energy-momentum tensor for the interacting system. It is of some 
interest to see how this field-theoretic approach can reproduce the above thermodynamic results.

Placing the interacting theory (\ref{LE1}) in a curved spacetime with metric $g_{\mu\nu}$, the
energy-momentum tensor is in general given by
\beq
     T_{\mu\nu}(x) = - {2\over \sqrt{g}}{\delta S[A]\over \delta g^{\mu\nu}(x)}      \label{tmunu}
\eeq
where $S[A]$ is the corresponding action functional for the system and $g = \det(g_{\mu\nu})$. 
The result in flat spacetime is given by the standard tensor
\beq
     T_{\mu\nu}^M = F_{\mu\lambda}F_{\nu\lambda} 
                  - {1\over 4} \delta_{\mu\nu}F_{\alpha\beta}F_{\alpha\beta}          \label{TMax}
\eeq
in free Maxwell theory plus a correction 
\beq
     T_{\mu\nu}^{EH} &=& {\alpha^2\over 45 m^4}
    \left(- 4F_{\mu\lambda}F_{\nu\lambda}(F_{\alpha\beta}F_{\alpha\beta})
      +  7F_{\mu\lambda}\tilde{F}_{\nu\lambda}(F_{\alpha\beta}\tilde{F}_{\alpha\beta})\right.\nn \\
     &+& \left.{1\over 2}\delta_{\mu\nu}\left[(F_{\alpha\beta}F_{\alpha\beta})^2 
              - {7\over 4}(F_{\alpha\beta}\tilde{F}_{\alpha\beta})^2\right]\right) \label{TEH}
\eeq
due to the Euler-Heisenberg interaction. 

\begin{figure}[htb]
 \begin{center}
  \epsfig{figure=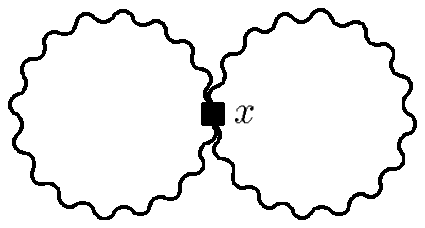, height=15mm}
 \end{center}
 \vspace{-8mm}            
 \caption{\small Feynman diagram giving the expectation value in the free theory of the 
interacting energy-momentum tensor}
 \label{figx}
\end{figure}

\begin{figure}[htb]
 \begin{center}
  \epsfig{figure=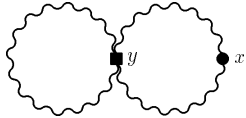, height=15mm}
 \end{center}
 \vspace{-8mm}            
 \caption{\small Feynman diagram for the non-local contribution to the expectation value of the 
          energy-momentum tensor}
 \label{figy}
\end{figure}

In the following it will be convenient to express the energy-momentum tensor $T_{\mu\nu} =
T_{\mu\nu}^M +  T_{\mu\nu}^{EH}$ in terms of the components of the electric and magnetic
fields. In particular, for the calculation of the energy density ${\cal E}  =  
\ex{T_{44}}_{int}$ where the expectation value is taken in the interacting theory, we
need the Maxwell result
\beq
    T_{44}^M = {1\over 2}\left({\bf E}^2 - {\bf B}^2\right)                     \label{T44}
\eeq
and the Euler-Heisenberg contribution
\beq
     T_{44}^{EH}= {2\alpha^2\over 45 m^4}\left({\bf B}^4 - 3{\bf E}^4 + 7({\bf E}\cdot{\bf B})^2
                - 2{\bf E}^2{\bf B}^2\right)                                    \label{EH44}
\eeq
Using the previous free equal-time correlators $\ex{E_iE_j} = -\ex{B_iB_j} 
= -\delta_{ij}\pi^2T^4/45$ 
and $\ex{E_iB_j} = 0$ combined with Wick's theorem, we obtain the correction 
\beq
     \ex{T_{44}^{EH}} = - {22\,\pi^4\alpha^2 \over 3^5\cdot 5^3}{T^8\over m^4}      \label{EEH1}
\eeq
due to the photon interactions. The result corresponds to the evaluation of the Feynman 
diagram in Fig.\ref{figx} which in the previous section appeared in the calculation of the 
correction to the free energy. 
But there is an additional contribution coming from the photon interactions as is seen from the 
general expression for the expectation value
\beq
     \ex{T_{44}}_{int} &=& {1\over Z}\int\!{\cal D}A_\mu \,(T_{44}^M + T_{44}^{EH})\,  
                       e^{-\int_0^\beta\! d^4x {\cal L}_E} \label{ex1} \\
              & = &\ex{T_{44}^M +T_{44}^{EH}    
               -  T_{44}^M\int_0^\beta\! d^4x \,\Delta{\cal L}_E}    \label{ex2} \\
              & \equiv & \ex{T_{44}^M} + \ex{T_{44}^{EH}} + \Delta\ex{T_{44}^M} \label{ex3}
\eeq
when expanded to lowest order in perturbation theory. Writing out the last
term, it is seen to be given by
\beq
    \Delta\ex{T_{44}^M}  &=&  {2\alpha^2\over 45 m^4}
    \int_0^\beta\! d^4y \cdot {1\over 2}\ex{[{\bf E}^2(x) - {\bf B}^2(x)]  \nn \\
  &\cdot &\left([{\bf E}^2(y) + {\bf B}^2(y)]^2  -  7({\bf E}(y)\cdot{\bf B}(y))^2\right)} 
                                               \label{DT44}      
\eeq 
It is a non-local contribution receiving contributions from all over spacetime and corresponds 
to the Feynman diagram in Fig.\ref{figy} with three propagators. 
This makes now the calculation more difficult than in the previous case of the 2-loop diagram in
Fig.\ref{figx}. 
With most of the needed details of the calculation in Appendix B, we find the result 
\beq
     \Delta\ex{T_{44}} &=& {\pi^4\alpha^2\over 3^4\cdot 5^3}{T^8\over m^4}
  \left[\left(-{10\over 3}\right) + 2\cdot 4 + (-10) - 7\cdot\left(+{4\over 3}\right)\right.\nn  \\
    &-& \left. 10 - 2\cdot (-8) - {50\over 3} + 7\cdot\left(-{8\over 3}\right)\right]
    = - {6\cdot 22\,\pi^4\alpha^2 \over 3^5\cdot 5^3}{T^8\over m^4}             \label{t44}
\eeq
Adding this contribution to (\ref{EEH1}) we obtain the previous thermodynamic result (\ref{E})
for the correction $\Delta{\cal E}$ to the Stefan-Boltzmann result.

While the energy-momentum tensor (\ref{TMax}) for the free Maxwell field is traceless, implying 
the relation ${\cal E} = 3P$ between energy density and pressure, the Euler-Heisenberg 
interaction $\Delta{\cal L}_E$ gives a contribution (\ref{TEH}) which has a non-zero trace. 
Actually, we now find 
\beq
        T_{\mu\mu} = 4\Delta{\cal L}_E                                    \label{trace}
\eeq
Since the correction to the pressure due to the interaction is $\Delta P = - \ex{\Delta{\cal L}_E}$,
we then have the relation ${\cal E} - 3P = 4\Delta P$  which is just the previous thermodynamic 
result $\Delta {\cal E} = 7\Delta P$.
 
\subsection{Response functions}
\begin{figure}[htb]
 \begin{center}
  \epsfig{figure=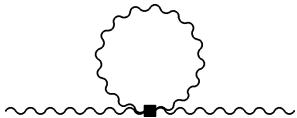, height=15mm}
 \end{center}
 \vspace{-8mm}            
 \caption{\small Euler-Heisenberg correction to the photon propagator.}
 \label{fig5}
\end{figure}

The polarization of the vacuum is described by the photon selfenergy $\Pi_{\mu\nu}(k)$. In lowest
order perturbation theory  it is given by the correlator
$\ex{A_\mu(k)\Delta{\cal L}_{E}(0)A_\nu(-k)}$ corresponding to the 1-loop Feynman diagram in 
Fig.\ref{fig5}.
 At zero temperature
it has been calculated by Halter\cite{Halter} who showed that the photon remains massless as
expected. Here it is correspondingly given by the sum-integral
\beq
    \Pi_{\mu\nu}(k) = {44\alpha^2\over 45m^4}\;\sumint_q \frac1{q^2}\left[(k\cdot q)^2\delta_{\mu\nu} 
     + k^2 q_\mu q_\nu - (k\cdot q)(k_\mu q_\nu + q_\mu k_\nu)\right]                  \label{PI}
\eeq
All the terms can be evaluated from the basic sum-integral (\ref{sumint}) which gives 
\beq
  \Pi_{\mu\nu}(k) &=& {44 \pi^2\alpha^2\over 2025}{T^4\over m^4}
  \left[k^2(\delta_{\mu\nu}  -  2\delta_{\mu 4}\,\delta_{\nu 4})\right.  \nn \\
  &-&\left. k_\mu k_\nu - 2k_4^2\delta_{\mu\nu} + 2k_4(k_\mu\delta_{\nu 4} 
                  + k_\nu\delta_{\mu 4})\right]         \label{PImunu}
\eeq
Introducing the transverse projection tensors $P_{\mu\nu}(k)$ and $Q_{\mu\nu}(k)$ as defined by
Weldon \\ \cite{Weldon} we can write the above result as
\beq
     \Pi_{\mu\nu}(k) = \Pi_T(k)P_{\mu\nu} + \Pi_L(k)Q_{\mu\nu}
\eeq
in terms of the scalar response functions $\Pi_T(k)$ and $\Pi_L(k)$. From the explicit form of the 
projection tensors, we then have in general
\beq
    \Pi_L(k) = - {k^2\over {\bf k}^2}\Pi_{44}, \hspace{10mm}  
    \Pi_T(k) = {1\over 2}(\Pi_{\mu\mu} + \Pi_L)                                       \label{Pis}
\eeq
From our result (\ref{PImunu}) we then get for these functions
\beq
      \Pi_L(k) = {44 \pi^2\alpha^2\over 2025}{T^4\over m^4}({\bf k}^2 + k_4^2)
\eeq
and
\beq
      \Pi_T(k) = {44 \pi^2\alpha^2\over 2025}{T^4\over m^4}({\bf k}^2 - k_4^2)
\eeq
The transition back to Minkowski space is now achieved by the standard analytical continuation
$ik_4 \ra \omega$ for the time component of the momentum vector while
the space components remain unchanged. 
In this way we obtain the physical response functions $\pi_{T,L}(\omega,\bk)$
from the above results. We then have the photon propagator in Minkowski space from which we can read
off the frequency dependent permittivity $\epsilon(\omega,\bk)$ and susceptibility $\mu(\omega,\bk)$.
One then finds\cite{Weldon}
\beq
        \epsilon(\omega,\bk) = 1 - {\pi_L\over k^2}
\eeq
where now $k^\mu = (\omega, \bk)$ with $k^2 = \omega^2 - \bk^2$ is the physical 4-momentum in
Minkowski space. Similarly,
\beq
    {1\over \mu(\omega,\bk)} = 1 + {k^2\pi_T - \omega^2\pi_L \over \bk^2 k^2}
\eeq
In the static limit $\omega \ra 0$ we thus obtain for the interacting photon gas
\beq
     \epsilon = \mu = 1 + {44 \pi^2\alpha^2\over 2025}{T^4\over m^4}
\eeq
so that the speed of light becomes
\beq
     c = \sqrt{1\over \epsilon\mu} = 1 - {44 \pi^2\alpha^2\over 2025}{T^4\over m^4}       \label{c}
\eeq
It is reduced below its ordinary vacuum value due to the interaction of the photons with virtual
electron-positron pairs as first calculated by Tarrach\cite{Tarrach}. A corrected result has since
then been derived in full QED from 2-loop diagrams\cite{LaTorre} and agrees with the above result
obtained more directly in the effective theory.

\section{The Casimir Effect}
In the previous section where the photons were at equilibrium at finite temperature, the 
corresponding electromagnetic field was periodic in the imaginary time dimension. It gives rise
to a finite, temperature-dependent vacuum energy. This situation is mathematically identical
to compactifying one of the spatial dimensions instead. For instance, one can imagine the field
at zero temperature confined by boundary conditions in such a way that it is periodic in the 
$z$-direction with period $L$. The result for the free energy calculated in the previous section
can then be used to obtain the vacuum energy for this confined field by the substitution
$T \ra 1/L$. In the case of free photons, the Stefan-Boltzmann free energy density (\ref{Free})
gives the vacuum energy density ${\cal E} = -\pi^2/45L^4$. This is now a quantum effect 
due to the vacuum fluctuations of the field. 

\subsection{Electromagnetic multipoles and quantization}
Vacuum energies induced by non-trivial boundary conditions in space imposed on a 
quantum field is now generically called Casimir energies and have been much 
studied\cite{Russian}\cite{Milonni}. 
We will here consider the original Casimir effect\cite{Casimir} due to fluctuations of the 
electromagnetic field $A^{\mu}(x) = (A_0(\bx,t),{\bf A}(\bx,t))$ confined between two parallel 
plates with separation $L$ and with normal ${\bf n}$ along the $z$-axis. We take the plates to be 
perfect conductors so that the fields satisfy the metallic boundary conditions 
\beq
     {\bf n}\times {\bf E} = {\bf n}\cdot {\bf B} = 0                            \label{bc}
\eeq
at all frequencies. For this problem we find it most convenient to work
in the Coulomb gauge ${\bg\nabla}\cdot{\bf A} = 0$ so that the Maxwell equation 
$\del_{\mu}F^{\mu\nu} = 0$ separates into the wave equation
\beq
    (\del_t^2 - \nabla^2){\bf A}(\bx,t) = 0                                 \label{wave}
\eeq
for the magnetic potential and the Poisson equation $\nabla^2 A_0(\bx,t)=0$ for the electric 
potential. Since the latter is zero on both plates, we can then simply take $A_0 = 0$. Thus we 
have the field strengths ${\bf E} = -\dot{\bf A}$ and ${\bf B} = {\bg\nabla}\times{\bf A}$.

The magnetic potential ${\bf A}(\bx,t)$ is most easily quantized by expanding it into 
electromagnetic multipoles\cite{LR}. From the symmetry of the problem we know that the solutions
of the wave equation (\ref{wave}) with a definite frequency $\omega$ will be labeled by wave 
vectors $\bk = (\bk_T, k)$ with $\omega = |\bk|$. Here $\bk_T = (k_x, k_y)$ represents the 
transverse components and $k = k_z$ is the longitudinal component. Transverse electric (TE) 
modes  will have $E_z = 0$ and can thus be written on the form
\beq
    {\bf A}^E_\bk({\bf x})  = {\bg\nabla}\times{\bf U}^E_\bk({\bf x})  \label{TE} 
\eeq
while the transverse magnetic (TM) modes have $B_z = 0$ so that 
\beq
    {\bf A}^M_\bk({\bf x})  = {\bg\nabla}\times
                             [{\bg\nabla}\times{\bf U}^M_\bk({\bf x})]       \label{TM}
\eeq
Using the wave equation (\ref{wave}) we have for the corresponding potentials 
\beq
    {\bf U}^E_\bk({\bf x})& = & N\,{\bf n} \sin{(k z)}\,e^{i{\bk_T}\cdot{\bx_T}}
                                                                                \label{UE}\\
    {\bf U}^M_\bk({\bf x}) & = & {N\over i\omega}\,{\bf n} 
    \cos{(k z)}\,e^{i{\bk_T}\cdot{\bx_T}}                                          \label{UM}
\eeq
where $\bx_T = (x,y)$. The boundary conditions (\ref{bc}) are now satisfied when taking
\beq
     k = {\pi\over L}n, \hspace{10mm} n=0,1,2,\ldots                               \label{k}
\eeq
The corresponding frequencies are therefore also quantized,
\beq
     \omega = (\bk_T^2 + (n\pi/L)^2)^{1/2}
\eeq
We see that the special mode with $n=0$ has $B_z = 0$ and is thus a TM mode. 
In this way we have found all the multipoles except for their absolute size given by
the constant $N$.  It can be determined from the normalization condition which we take to be
\beq
    \int\!d^3x\,{\bf A}^\lambda_{n \bk_T}(\bx)\cdot
                {\bf A}^{\lambda'}_{n'\bk'_T}(\bx) 
    = \delta_{\lambda\lambda'}\,\delta_{nn'}\,(2\pi)^2\delta(\bk_T - \bk'_T)
                                                                                  \label{norm}
\eeq
where $\lambda = (E,M)$. After some integrations, one obtains
\beq
     N = \left\{\begin{array}{ll}
                      {1\over k_T}\sqrt{1\over L}, & \mbox{$ n = 0 $} \\
                      {1\over k_T}\sqrt{2\over L}, & \mbox{$ n = 1,2,\ldots $}
                      \end{array}\right.                                           \label{N}
\eeq    
It is the $n=0$ TM mode  which has the exceptional normalization.
The full vector potential can now be expanded in terms of these transverse modes as
\beq
     {\bf A}(\bx,t) = \sum_{\lambda=E,M}\sum_{n=0}^\infty
                      \int\!{d^2k_T\over(2\pi)^2}\sqrt{1\over2\omega}
    \left[a^\lambda_{n \bk_T}{\bf A}^\lambda_{n \bk_T}(\bx)e^{-i\omega t}
    + c.c.\right]                                                               \label{modes}
\eeq
In the quantum theory the expansion coefficients $a^\lambda_{n \bk_T}$ are annihilation 
operators satisfying the canonical commutator
\beq
     [a^\lambda_{n \bk_T},a^{\lambda'\dagger}_{n'\bk'_T}] =
     \delta_{\lambda\lambda'}\,\delta_{nn'}\,(2\pi)^2\delta(\bk_T - \bk'_T)
                                                                              \label{quant}
\eeq
Since the vacuum state by definition is annihilated by these operators, the expectation 
values of the electric and magnetic fields in this state are zero. But the fields have non-zero
fluctuations which will be calculated. 

\subsection{Casimir energy}
The most important physical effect of the vacuum fluctuations will be the attraction
between the plates discovered by Casimir\cite{Casimir} and now experimentally verified\cite{exp}.
It is due to the zero-point energy of the vacuum state and is given by the divergent expression
\beq
     E = {1\over 2}\int\!{d^2k_T\over(2\pi)^2} \left[k_T + 2\sum_{n=1}^\infty
           \left(\bk_T^2 + ({n\pi\over L})^2\right)^{1\over 2}\right]   \label{E_0}
\eeq
The first term is due to the TM $n=0$ mode while the sum includes equal contributions from the 
TE and the TM modes. Obviously, both the sum and the integrals are strongly divergent. 
Usually they are
regulated by a UV cutoff\cite{Milonni}, but we will find it more convenient to use dimensional
regularization of the transverse momentum integrations combined with zeta-function 
regularization of the longitudinal momentum summations as in the previous section. We will 
therefore again use the standard integral
\beq
     I(d,p) =  \int\!{d^d k_T\over(2\pi)^d}\left(\bk_T^2 + k^2\right)^{-{p\over 2}}  
         = {k^{d-p}\over (4\pi)^{d/2}}{\Gamma({p-d\over 2})\over\Gamma({p\over 2})}  \label{int}
\eeq
When $k=0$ the result is zero so that the first term in (\ref{E_0}) gives zero. In the second 
term we need $I(2,-1) = -k^3/6\pi$ so that the Casimir energy becomes
\beq
     E = -{1\over 6\pi}\sum_{n=1}^\infty\left({\pi\over L}n\right)^3 
         = - {\pi^2\over 720L^3}                                                     \label{E0}
\eeq
when we use the value $\zeta(-3) = 1/120$ for the Riemann zeta-function. This is the energy per
unit plate area so that the energy density is  ${\cal E} = E/L = -\pi^2/720L^4$. The attractive
Casimir force between the plates, is given by the derivative of the energy (\ref{E0}) and thus
becomes $F = - \pi^2/240L^4$. We notice that the above energy density follows from the 
Stefan-Boltzmann free energy density (\ref{Free}) by the 
substitution $T\ra 1/2L$. Forcing the field to satisfy metallic boundary conditions therefore
corresponds to taking $L$ to be half of the period $1/T$ at finite temperature.

\subsection{Field fluctuations}
In the above we have implicitly assumed that the energy density is constant between the two 
plates. But there is no deep reason for that to be the case. For instance, the Casimir energy
density inside a spherical, metallic shell increases outward from the center and diverges when 
one approaches the shell\cite{OR}. In the present case with parallel plates the 
energy density is in fact constant. However, as soon as we include field interactions we will
find that the density varies with the position between plates.

The electric field ${\bf E} = ({\bf E}_T, E_z)$ is given by the time derivative of the magnetic 
potential (\ref{modes}). For the fluctuations in the transverse components we then find
\beq
   \ex{{\bf E}_T^2(\bx)} =
   {1\over L}\sum_{n=1}^\infty  \int\!{d^2k_T\over(2\pi)^2}
   \left(\omega + {k^2\over\omega}\right)\sin^2{(kz)}                          \label{ETET}
\eeq
while the longitudinal fluctuations are given by
\beq
   \ex{{E}_z^2(\bx)} = 
   {1\over L}\sum_{n=1}^\infty  \int\!{d^2k_T\over(2\pi)^2}
   \left(\omega - {k^2\over\omega}\right)\cos^2{(kz)}                         \label{EZEZ}
\eeq
Corresponding expressions for the fluctuations in the magnetic field are also easy to derive
and can in fact be directly obtained from the relations $\ex{B_x^2} = \ex{B_y^2}
= - \ex{E_z^2}$ and $\ex{B_z^2} = - \ex{E_x^2}$. Using the regularized integral (\ref{int})
which gives $I(2,1) = -k/2\pi$, we find that the transverse fluctuations become
\beq
     \ex{{\bf E}_T^2(\bx)} 
      = - {\pi^2\over 3L^4}\sum_{n=1}^\infty n^3 (1 - \cos{2\theta n})         \label{ET}
\eeq    
where $\theta = \pi z/L$ is the dimensionless distance between the plates. Again with
zeta-function regularization the first sum is $\zeta(-3) = 1/120$ while the last sum becomes
\beq
    \sum_{n=1}^\infty n^3 \cos{2\theta n} = -\left({1\over 2}{d\over d\theta}\right)^3
    \sum_{n=1}^\infty \sin{2\theta n}  = {1\over 8}F(\theta)                    \label{sum}
\eeq
where the function
\beq
     F(\theta) = -{1\over 2}{d^3\over d\theta^3}\cot{\theta} 
               = {3\over\sin^4{\theta}} - {2\over\sin^2{\theta}}                 \label{F}
\eeq
gives the position-dependence of the field fluctuations,
\beq
    \ex{{\bf E}_T^2(\bx)}^{reg} 
                 = - {\pi^2\over 24L^4}\left({1\over 15} - F(\theta)\right) \label{ETx}
\eeq
Similarly, we obtain for the longitudinal fluctuation (\ref{EZEZ}) 
\beq
   \ex{{E}_z^2(\bx)} = {\pi^2\over 6L^4}\sum_{n=1}^\infty n^3 (1 + \cos{2\theta n}) \label{Ez}
\eeq
which then becomees
\beq
    \ex{{E}_z^2(\bx)}^{reg} = {\pi^2\over 48L^4}\left({1\over 15} + F(\theta)\right) \label{Ezx}
\eeq
after regularization. These results now also describe the fluctuations of the magnetic field using 
the above relations to the corresponding electric fluctuations. Combining the transverse and 
longitudinal components we thus have for the fluctuations of the full field strengths
\beq
     \ex{{\bf E}^2(\bx)}^{reg} 
             = -{\pi^2\over 16L^4}\left({1\over 45} - F(\theta)\right), \hspace{10mm}
     \ex{{\bf B}^2(\bx)}^{reg} 
             = -{\pi^2\over 16L^4}\left({1\over 45} + F(\theta)\right)   \label{plates}
\eeq
In addition we have $\ex{E_iB_j} = 0$ which also follows directly from time-reversal invariance.

These results have previously been obtained by L\"utken and Ravndal\cite{LR} using a different
regularization scheme formed by letting the difference between the time coordinates of the two
fields in the correlators be imaginary and slightly non-zero in magnitude. More recently,
Bordag {\it et al} \cite{Bordag} have considered the same problem in the Lorentz gauge and 
obtained a closed expression for the photon propagator in the presence of the two plates. Letting
the two coordinates in the propagator approach each other, one obtains expressions for the local
fluctuations of the quantum fields. This is done in Appendix C. After regularization, now based 
upon a point-split in the spatial direction, we then again recover the above results.

Because of the boundary condition (\ref{bc})  the transverse electric field is constrained to be 
zero on the walls. The fluctuations in the transverse components must therefore be zero for
$z = 0$ and $z=L$ which we see is consistent with the general expression (\ref{ETET}). However, the
regularized fluctuations (\ref{plates}) are seen to diverge when one approaches one of the 
plates. When $z\ra 0$ the leading terms are
\beq
     \ex{{\bf E}^2(\bx)}_{z\ra 0}^{reg} =  {3\over 16 \pi^2 z^4},             \hspace{10mm} 
     \ex{{\bf B}^2(\bx)}_{z\ra 0}^{reg} = - {3\over 16 \pi^2 z^4}             \label{plane}
\eeq
These expressions will represent the field close to any metallic surface, plane or curved,
since when one gets close enough, it will always appear as plane. Since the
transverse fluctuations are exactly zero on the plate and highly divergent just outside, it
is clear that the physics very close to the plate is complicated. In fact, it will depend
on the microscopic properties of the wall reflected in the existence of a non-zero cut-off 
in this region.  In other words, when one gets very close to a wall, the detailed properties
of the fluctuations will depend on this cut-off and the regularization becomes dependent on
which scheme is being used. Using for example a non-zero point-split $\e$ in the imaginary time
dimension, it is easy to show\cite{KR} that the limit $\e \ra 0$ cannot be taken when the 
distance to the wall is less than $\e$. And the magnitude of $\e$ in our effective field 
theory approach is set by the inverse electron mass.

The position-dependence of the correlators $\ex{{\bf E}^2}$ and $\ex{{\bf B}^2}$ can in principle
be measured. An atom outside a single plate will be attracted to the plate by
the Casimir-Polder force\cite{CP} which has a position dependence directly given by the one 
in the correlators\cite{LR}. This force, or its equivalent van der Waals force at shorter
distances, has now been experimentally verified\cite{CPex}. The electric fluctuations 
$\ex{{\bf E}^2}$ will affect the energy levels of an atom near a wall\cite{LR}. By moving the 
atom around one can then map the spatial variation of the vacuum fluctuations by 
spectroscopic methods\cite{QED}. One will then observe that these effects become much stronger
as one gets very near the plate or wall. But the obtained results for the local fluctuations
are not valid when one gets too close to a wall. Then new physics determined by microscopic 
properties of the wall will start to become relevant and are outside the present description.

\subsection{Euler-Heisenberg interactions}
It is of interest to calculate the radiative corrections to the one-loop Casimir energy 
(\ref{E0}) due to the interactions of the fluctuating electromagnetic field with the 
virtual electrons in the vacuum. The first correction in QED comes from the 
Uehling interaction in Fig.\ref{fig3} which has been evaluated by Xue\cite{Xue}. 
Instead of the physical
boundary conditions (\ref{bc}) he constrained the photon field to be periodic in the 
$z$-direction with period $2L$. As previously noted, this gives the correct result for the
1-loop term. He finds that the 2-loop contribution is exponentially suppressed like 
$\exp{(-2mL)}$ and therefore completely negligible in the physical limit $L \gg 1/m$.
Since the system with these boundary conditions is now mathematically equivalent to being
at finite temperature $T = 1/2L$, this outcome is consistent with  result of 
Woloshyn\cite{Woloshyn} discussed in the first section.

A much different calculation by Bordag {\it et al} \cite{Bordag} and later by Robaschik {\it et al}
\cite{Robaschik} obtains a radiative
correction to the Casimir force proportional with $\alpha/mL^5$ using the physical boundary 
conditions (\ref{bc})  for the photon field. The electron field is assumed to be 
unaffected by the metallic plates.
The correction is small and decreases faster with the plate separation than the leading term. 
But it is somewhat surprising to see that it depends on an odd power of the electron mass. 
Since this calculation is done within full QED, it should be valid down to distances of the 
order of the inverse electron mass. So in this case one is no longer discussing a physical
plate made up of atoms, but a more ideal, mathematical situation. It is thus not obvious
that  this result should agree with what is obtained using low-energy, effective field theory.
Then the first radiative correction to the 1-loop result should result from the Uehling term 
in (\ref{LU}). Since it is proportional to $\alpha/m^2$, it could give a correction to the 
Casimir effect varying like $\alpha/m^2L^6$ for dimensional reasons. However, since the 
Uehling interaction can be transformed away, the first non-zero contribution will result from
the Euler-Heisenberg interaction (\ref{LEH}). In Minkowski space it is given by the perturbation
\beq
     \Delta{\cal L} = {2\alpha^2\over 45 m^4}\left[({\bf E}^2 - {\bf B}^2)^2 
                  + 7 ({\bf E}\cdot{\bf B})^2\right]                           \label{MEH2}
\eeq
In analogy with what we did at finite temperature in the previous section, we now can obtain
the correction to the Casimir energy from
\beq
    \Delta E = - \int\!d^3x \ex{\Delta {\cal L}}                              \label{DE}
\eeq
in lowest order perturbation theory. 

Let us first assume that this energy correction can be written as an integral over a local 
energy density $\Delta\hat{\cal E}$. Its magnitude will be given by $\Delta\hat{\cal E} = - 
\ex{\Delta{\cal L}}$. Using Wick's theorem in the form $\ex{E_i^2E_j^2} = \ex{E_i^2}\ex{E_j^2} 
+ 2\ex{E_i^2}^2\delta_{ij}$ with the expectation  values from (\ref{ETx}) and (\ref{Ezx}) we 
then obtain 
\beq
    \Delta\hat{\cal E} &=& -{2\alpha^2\over 15m^4}\left[5\ex{E_x^2}^2 + 5\ex{E_z^2}^2 
                       - \ex{E_x^2}\ex{E_z^2}\right]                                   \nn \\
    &=&-{\alpha^2 \pi^4\over 2^73^35m^4L^8}\left[{11\over 225}+9F^2(\theta)\right] \label{deltau}
\eeq              
Because of the small, numerical coefficient in front and the condition $L\gg 1/m$ this 
correction is seen to be negligible for all practical purposes. However, its magnitude gets 
larger as one approaches one of the plates where it is seen to diverge like
$\Delta\hat{\cal E}(z\ra 0) = -\alpha^2 \pi^4/ 1620m^4z^8$. The integral of this energy 
density  over the space between the plates, will thus also diverge. Such a result is meaningless 
and a forces us to question the physical content of this position-dependent energy density. 

If we instead first do the volume integral in (\ref{DE}) and afterwards regularize the 
result, we get a finite result for the Casimir energy correction. It results from terms with the
typical behavior of
\be
   \int_0^L dz \ex{E_z^4(\bx)} &=& 3\int_0^L dz \ex{E_z^2(\bx)}\ex{E_z^2(\bx)}\\
   &=& 3\left({\pi^2\over 6L^4}\right)^2\int_0^L dz\!\!\sum_{m,n = 1}^\infty m^3n^3
   [1 + \cos{(2\pi zm/L)}][1 + \cos{(2\pi zn/L)}]
\ee
The constant in the integrand simply gives $\zeta^2(-3) = (1/120)^2$ while the integrals over
the cosines vanishes because of the boundary conditions. The last integral, from the product of 
the two cosines, is only non-zero when $m=n$. But then its value will be multiplied by 
$\zeta(-6) = 0$ and is thus also zero.  So it is only the position-independent parts of the
field fluctuations that contribute to the total energy. The net result for the integrated 
correction to the vacuum energy is therefore given by just the first term in (\ref{deltau}), 
i.e. 
\beq
    \Delta E = - {11\alpha^2\pi^4\over 2^7\cdot 3^5\cdot 5^3 \, m^4L^7}    \label{DE0}
\eeq
Dividing by the plate separation $L$ to get an energy density, we see that it can also be 
obtained from the last term in the free energy (\ref{F2}) by the substitution $T \ra 1/2L$ just 
as the leading term of the Casimir energy could. It represents a tiny, additional contribution 
to the attractive force between the plates. 

The lack of commutativity between integration and regularization we have just observed in the 
calculation of the vacuum energy, is already present in the case of free fields as 
first pointed out by Deutsch and Candelas\cite{DC}. It can be seen by integrating for example
the electric field fluctuations (\ref{ET}) and (\ref{Ez}) over the separation between the plates,
which gives the electric half of the  Casimir energy (\ref{E0}) after regularization. Only the
position-independent portion of the fluctuation contribute to the result. On the other hand,
integrating first the regularized field fluctuation in (\ref{plates}) obviously gives a divergent
result. The regularization dampens out the effect of the high energy modes of the 
field which  prevents us from making any reliable predictions near the plates. As a 
consequence, one is not allowed to integrate a regularized result over the full volume between 
the plates. As shown by Scharnhorst\cite{Scharnhorst} and Barton\cite{Barton_1} the same vacuum 
fluctuations will also modify the speed of light in the space between the plates. But again one 
finds that the position-dependence does not contribute to the net effect.

An unambiguous result for the energy density and pressure between the plates can be obtained
from the energy-momentum tensors (\ref{TMax}) and (\ref{TEH}) in the same way as was done in the
previous section for the photon gas at finite temperature. In the expectation value of the free 
Maxwell part the position-dependent terms again cancel out and gives the traceless result
\beq
    \ex{T_{\mu\nu}^M} = {\pi^2\over 720 L^4}(\eta_{\mu\nu} - 4n_\mu n_\nu)          \label{tmn}
\eeq
Here the four-vector $n^\mu = (0;0,0,1)$ represents the normal ${\bf n}$ to the plates and we 
are using a metric $\eta_{\mu\nu}$ with positive signature. The energy density ${\cal E} = 
\ex{T_{00}}$ is obviously consistent with the integrated result (\ref{E0}) while $\ex{T_{zz}}$
gives the Casimir force. Similarly, the contribution $\ex{T_{\mu\nu}^{EH}}$ from the 
interactions is fairly straightforward to calculate using the same field correlators as 
above. However, the evaluation of  $\Delta\ex{T_{\mu\nu}^M}$ in (\ref{DT44}) is now much more 
difficult than at finite temperature. One must use the full field propagator discussed in
Appendix C where the transverse and longitudinal components behave differently. But the main
reason for the complications is the lack of momentum conservation in the $z$-direction which 
results in highly divergent, oscillating double sums. The evaluation of these will require a
separate investigation. At this stage we can only say that the energy-momentum tensor for
the interacting system must satisfy the trace condition $T^{\mu}_{\mu} = - 4\Delta{\cal L}$ which
follows from (\ref{trace}) when taken to Minkowski space. The expectation
value of the right hand side is here the energy density $\Delta\hat{\cal E}$ in (\ref{deltau}).

\section{Discussion and conclusion}
Effective field theory is a very powerful framework for calculating higher order quantum effects
of interacting fields below characteristic energy scale. In our case of low-energy photons this is
set by the electron mass. The Stefan-Boltzmann energy increases with temperature like $T^4$ and one
would expect for dimensional reasons that the first radiative correction would go like $T^6/m^2$.
We find that this term is in fact absent at low temperatures since the Uehling interaction can 
be transformed away in the absence of matter. The first correction comes in at
next order and is due to the Euler-Heisenberg effective interaction. It gives a contribution
proportional to $T^8/m^4$. In the effective theory it results essentially from a simple 1-loop
calculation, while in full QED it would result as the low-energy limit of a 3-loop calculation.
The absence of the $T^6$ term in the energy density is also observed for the interacting pion gas
calculated in the chiral limit of massless fields\cite{Gerber}. If there is any connection here,
the reason is not clear.

We have also calculated the finite-temperature polarization tensor of the interacting photons. In
the effective field theory approach it is obtained from a 1-loop calculation while in full QED it 
would come from a more difficult 2-loop calculation. The resulting permittivity and susceptibility
gives an effective light velocity which is less than one. At normal temperatures one can safely forget
the small correction, but it will increase for extremely high temperatures which one has in the early
universe. If it has any practical consequences in this context, does not seem to have  been 
investigated. But it is clear that the propagation of light over cosmological distances is 
affected by the finite-temperature quantum fluctuations in the vacuum.

Much of these finite temperature results can be applied directly to the Casimir effect induced
by two parallel plates held at zero temperature. The absence of the $T^6$ term in the free energy
implies that the first quantum correction to the Casimir energy density is  caused by the 
Euler-Heisenberg interaction and varies like $1/m^4L^8$ with the plate separation.
If the boundary conditions of the electromagnetic field on the plates could be taken to be periodic, 
one could in fact obtain the Casimir energy simply by letting $T \ra 1/L$ in the finite-temperature 
results. The field fluctuations $\ex{{\bf E}^2}$ and $\ex{{\bf B}^2}$ would be constant between the 
plates and there would be no divergences as one gets near one of the plates. 
But the physical boundary conditions we use forces some of the components of the fields to be 
exactly zero at the plates. This classical constraint on the fields induces large quantum 
fluctuations which in fact diverge close to the boundary. The corresponding electric and magnetic
energy densities diverge also and thus also their integrated contributions to the total energy. For
the free field, the divergences cancel so that one is left with a finite total Casimir energy. But
for interacting photons, the energy density is in general divergent near the plates. In spite of
this, we have obtained a finite Casimir energy for the interacting system by integrating the energy
density first and then regularizing the result. We see that he processes of integration and 
regularization do not in general commute with each other.

In order to analyze this apparently paradoxical situation a bit more carefully, let us look at the
divergent sum 
\beq
    S(\theta) = \sum_{n=1}^\infty \sin{2\theta n}
\eeq   
which enter the calculation of the basic correlator (\ref{ET}). The field fluctuations
are given by its third derivative resulting in the function $F(\theta)$ in (\ref{sum}) when we use
zeta-function regularization. When $\theta = 0$ or $\theta = \pi$ each term in the sum
is zero while the regularized sum is infinite. So it cannot be used on the plates. A more physical 
understanding  of the sum is obtained by evaluating it instead with an exponential cutoff as a 
regulator. Defining the convergent sum
\beq
   S(\e,\theta) = \sum_{n=1}^\infty e^{-\e n}\sin{2\theta n}
\eeq
we will then have $S(\theta) = \lim_{\e\ra 0}S(\e,\theta)$. The cutoff should go to zero in the 
limit where the electron mass becomes very large since it is $m$ which sets the lower scale
for the frequencies which should be dampened out of the problem. Since the regulated sum is a
geometric series, it is easily found to be
\beq
   S(\e,\theta) = {e^{-\e}\sin 2\theta\over 1 - 2e^{-\e}\cos 2\theta + e^{-2\e}}
\eeq
Expanding this result in powers of $\e$ one finds
\beq
   S(\e,\theta) = {1\over 2}\cot\theta - {1\over 8}\;{\cos\theta\over\sin^3\theta}\;\e^2   
                + {\cal O}(\e^4)                                              \label{expand}
\eeq
We see that as long as $\sin\theta > 0$ one can take the limit $\e\ra 0$ and the first term will
represent the fluctuation sum $S(\theta)$ as we have done in the text. However, near the plates 
where $\sin\theta \ra 0$
the higher order terms in (\ref{expand}) will become more and more important and the first term will
no longer alone represent the fluctuations. The higher order terms with finite $\e$ must then be 
kept and represent in some way counterterms which must be introduced when one approaches one of the 
plates.

In particular, this applies when we want to calculate the total energy between the plates. Then we
are faced with the integral
\be
   \int_0^\pi\!d\theta {d^3\over d\theta^3}S(\e,\theta) =
   \left[{d^2\over d\theta^2}
   \left({e^{-\e}\sin 2\theta \over 1 - 2e^{-\e}\cos 2\theta + e^{-2\e}}\right)\right]_0^\pi
\ee
On the right-hand side we must take the derivative while $\e$ is kept finite. The integral gives 
then
zero at the upper and lower limits. This is consistent with what we found in the text that the
integral of the position-dependent energy density is in fact zero and not infinite. But it obtains 
only when we keep all the higher order terms in (\ref{expand}) since we need the fluctuations on the
boundary when we do the integral.

In this way we get a finite result for the integrated Casimir energy also when photon interactions
are included. All the position-dependence from the correlators integrate out to zero and we are
left with a final expression which can be obtained simply by letting $T \ra 1/2L$ in the 
corresponding result for the finite-temperature free energy of the photon gas. 

We want to thank E. Braaten, L.S. Brown and K. Scharnhorst for useful discussions, suggestions and
clarifying comments. We thank NORDITA for their generous support and hospitality during the 
completion of this work.

Xinwei Kong has been supported by the Research Council of Norway. 

\section*{Appendix A}
In order to illustrate how equations of motions can be used in effective theories, let us consider 
 a scalar field $\phi$ coupled to fermions $\psi$ and described by the effective Lagrangian
\be
    {\cal L} = -{1\over 2}\phi\Box\phi - g\phi\bar{\psi}\psi + {\alpha\over M^2}(\Box\phi)^2 
\ee
Since we want to simulate effective Maxwell theory in the text, the scalar field is taken to be 
massless. 
\begin{figure}[htb]
 \begin{center}
  \epsfig{figure=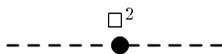}
 \end{center}
 \vspace{-8mm}            
 \caption{\small Derivative correction to the boson propagator.}
 \label{AppA.1}
\end{figure}
The last interaction term where $\alpha$ is some dimensionless coupling constant and 
$M$ is a heavy mass, is meant to be the corresponding Uehling interaction.  It gives a correction to
the scalar propagator shown in Fig.\ref{AppA.1}. 

The Yukawa coupling
$g$ gives rise to a fermion-fermion interaction in lowest order $g^2$ due to an exchange of
a massless $\phi$-particle. Because of the above Uehling term, there will be corrections to this
Yukawa interaction. 
\begin{figure}[htb]
 \begin{center}
  \epsfig{figure=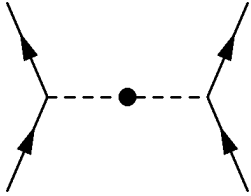}
 \end{center}
 \vspace{-8mm}            
 \caption{\small Propagator correction to the four-fermion interaction.}
 \label{AppA.2}
\end{figure}
To lowest order in $\alpha$ it results from the Feynman diagram in 
Fig.\ref{AppA.2} which gives the effective vertex
\be
  {\cal H}_{eff} &=&  g^2{\alpha\over M^2}\int\! d^4y \int\!{d^4k\over (2\pi)^4} 
                      \bar{\psi}(x)\psi(x)
                {1\over k^2} k^4 {1\over k^2} e^{ik\cdot(x-y)}\bar{\psi}(y)\psi(y) \\
                    &=& g^2{\alpha\over M^2}(\bar{\psi}(x)\psi(x))^2
\ee
which is a local four-fermion interaction. 

This result can also be obtained by using the equation of motion $\Box\phi = g\bar{\psi}\psi$ for
the scalar field which results from the dimension $D=4$ part of the Lagrangian. Putting now this 
back into ${\cal L}$, we obtain
\be
    {\cal L}' = -{1\over 2}\phi\Box\phi - g\phi\bar{\psi}\psi 
              + {\alpha\over M^2}g\bar{\psi}\psi\Box\phi
\ee
when we keep only interactions of dimensions $D=6$ or less. In this way we have now generated a new
derivative Yukawa coupling between the fermion and the scalar field. When combined with the original
Yukawa coupling, we obviously get the same, effective four-fermion interaction as above now due to
exchange of massless $\phi$ particle with the standard $1/k^2$ propagator.
Instead of using the equation of motion, we can perform the field redefinition
\be
   \phi \ra \phi + {\alpha\over M^2}\Box\phi
\ee
in the Lagrangian ${\cal L}$. Keeping only terms to order $1/M^2$, we see that we then recover the
modified Lagrangian ${\cal L}'$. 

\begin{figure}[htb]
 \begin{center}
  \epsfig{figure=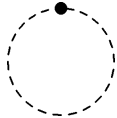,height=15mm}
 \end{center}
 \vspace{-8mm}            
 \caption{\small Lowest correction to the vacuum energy.}
 \label{AppA.3}
\end{figure}

When there is no fermionic matter present, the Uehling interaction can thus be transformed away. 
Hence, it should also then give no contribution to the vacuum energy. This can also be seen by
considering the lowest order Feynman diagram in Fig.\ref{AppA.3} resulting from the original 
Lagrangian ${\cal L}$. It gives a contribution 
\be
    \Delta{\cal E} = {\alpha\over M^2}\int\!{d^4k\over (2\pi)^4} {k^4\over k^2} = 0
\ee
when we use dimensional regularization. Had we instead used a momentum cutoff $\Lambda$, it would
have been $\propto\Lambda^6$ and would be cancelled by an appropriate counterterm. Also at non-zero 
temperature this contribution to the free energy can be set equal to zero.

If we denote
the scalar propagator by $D(x-y)$, the same diagram in coordinate space would have given
\be
    \Delta{\cal E} = \lim_{x\ra y} {\alpha\over M^2}\Box_x^2 D(x-y) 
                     = {\alpha\over M^2}\delta^{''}(0)
\ee
which obviously is the same as the above result. But the point now is that if the scalar field is
confined by some bounding surfaces, its propagator will instead be changed into $\bar{D}(x,y)$.
The Casimir energy density due to the the same diagram is then given by
\be
    \Delta{\cal E} = \lim_{x\ra y} {\alpha\over M^2}\Box_x^2 \bar{D}(x,y)
\ee
But even the confined propagator satisfy the basic equation $\Box_x \bar{D}(x,y) = \delta(x-y)$
so that the contribution to the vacuum energy density is the same as
for an unconfined  field and obtained 
above. The Uehling contribution to the Casimir energy is therefore zero.

\section*{Appendix B}
We will here present the electromagnetic field correlators at finite
temperature and derive some of the summation-integrals which appear in
evaluating the relevant Feynman diagrams.

From the basic, free field correlator (\ref{FF}) we can immediately write down
the corresponding electric propagators, 
\beq 
    \ex{E_i(x\,)E_j(y)} = \sumint_k {k_4^2 \delta_{ij} + k_ik_j\over \bk^2 + k_4^2}\; 
                          e^{ik\cdot (x - y)}                          \label{EiEj} 
\eeq 
the magnetic propagators, 
\beq 
     \ex{B_i(x)\,B_j(y)} = \sumint_k {\bk^2 \delta_{ij} - k_ik_j\over \bk^2 + k_4^2}\;
                            e^{ik\cdot (x - y)}                          \label{BiBj} 
\eeq 
and the mixed propagator 
\beq 
     \ex{E_i(x)\,B_j(y)} = \sumint_k {\epsilon_{ij\ell}\,k_4k_\ell\over \bk^2 + k_4^2}\; 
                            e^{ik\cdot (x -  y)}                           \label{EiBj} 
\eeq 
Notice that this latter one is zero when $x \ra y$ so that the expectation value 
$\ex{{\bf E}\cdot {\bf B}} = 0$ consistent with parity invariance.

The electric field fluctuation $\ex{{\bf E}^2}$ now follows from the propagator (\ref{EiEj})
by taking the coincidence limit $x \ra y$.  We then recover the result (\ref{EE}) in the 
main text. It can be evaluated from the basic integrals
\beq
     I_{ij} = \sumint_k {k_i\,k_j \over \bk^2 + k_4^2}
\eeq
and
\beq
     I_{44} = \sumint_k {k_4^2\over \bk^2 + k_4^2}
\eeq
both of which can be obtained from the master integral (\ref{int}) in the text. From rotational 
invariance we have that $I_{ij} = \delta_{ij}I_{kk}/3$ so we need only $I_{kk}$.
With dimensional regularization it follows that $I_{44} + I_{kk} = 0$ and one is left 
with only the integral $I_{44}$. Then it follows that
\be
    I_{44} = T \sum_{n= -\infty}^\infty \int\!{d^dk\over (2\pi)^d}\;
             {(2\pi nT)^2\over \bk^2 + (2\pi nT)^2} 
           = 2T \sum_{n=1}^\infty {(2\pi nT)^{d}\over (4\pi)^{d/2}}
             {\Gamma (1 - d/2)\over \Gamma(1)}
\ee
There are no singular poles in the limit $d \ra 3$ where we find
\be
    I_{44} = 2\,T^4 \pi^{3/2}\Gamma(-1/2) \sum_{n=1}^\infty n^3 = -\pi^2T^4/30
\ee
using zeta-function regularization for the divergent Matsubara sum with $\zeta(-3) = 1/120$. 
Together with the corresponding result for $I_{ij}$ we then have the basic integral 
(\ref{sumint}).

The full propagators enter the calculation of the extra term $\Delta\ex{T_{44}}$ in (\ref{DT44}).
Let us consider the very first term which can be expanded to give
\beq
    \int_0^\beta\! d^4y\,\ex{{\bf E}^2(x){\bf E}^4(y)} &=&
  4\int_0^\beta\! d^4y\,\ex{E_i(x)E_j(y)}\ex{E_i(x)E_j(y)}\ex{E_k(x)E_k(y)} \nn \\
    &+& 8\int_0^\beta\! d^4y\,\ex{E_i(x)E_j(y)}\ex{E_i(x)E_k(y)}\ex{E_j(x)E_k(y)}     \label{E6}
\eeq
Now the last factor in the first term is just $\ex{{\bf E}^2}$ which we already know. 
Integrating over $y$ makes the momenta in the two first propagators in the same term equal and 
we have
\beq
    \int_0^\beta\! d^4y\, \ex{E_i(x)E_j(y)}\ex{E_i(x)E_j(y)}\ex{E_k(x)E_k(y)} 
= \sumint_k \left({k_4^2\delta_{ij} + k_ik_j\over \bk^2 + k_4^2}\right)^2\ex{{\bf E}^2} \label{E6.x}
\eeq
Here we need several new integrals which again can be calculated with the use of (\ref{int}).
The first one gives
\be
     J_{44} = \sumint_k {k_4^4\over (\bk^2 + k_4^2)^2} 
            = 2T \sum_{n=1}^\infty {(2\pi nT)^3\over (4\pi)^{3/2}}
             {\Gamma (1/2)\over \Gamma(2)} = \pi^2T^4/60
\ee
when done like $I_{44}$ above. Now the integral
\beq
   J_4 = \sumint_k {k_4^2\bk^2\over (\bk^2 + k_4^2)^2}
\eeq
follows from the identity $J_{44} = I_{44} - J_4$ which gives $J_4 = -\pi^2T^4/20$. Similarly, 
the integral
\beq
    J =  \sumint_k {\bk^4\over (\bk^2 + k_4^2)^2}
\eeq
can be obtained from $J_4 = I_{kk} - J$ which gives $J = \pi^2T^4/12$. Using these results in 
(\ref{E6.x}), we then have 
\beq
    \int_0^\beta\! d^4y\, \ex{E_i(x)E_j(y)}\ex{E_i(x)E_j(y)}\ex{E_k(x)E_k(y)} 
    = - {\pi^4T^8\over 2\cdot 3^2\cdot 5^2}
\eeq
for the first term in (\ref{E6}).    

The second term can now also be obtained using  the same integrals,
\beq
     && \int_0^\beta\! d^4y\,\ex{E_i(x)E_j(y)}\ex{E_i(x)E_k(y)}\ex{E_j(x)E_k(y)} \\
   &=& \sumint_k {(k_4^2 \delta_{ij} + k_ik_j)(k_4^2 \delta_{ik} + k_ik_k)\over (\bk^2 + k_4^2)^2}
      \cdot \sumint_p {p_4^2\delta_{jk} + p_jp_k\over \bp^2 + p_4^2} 
  = -  {\pi^4T^8\over 2\cdot 3^3\cdot 5^2}
\eeq 
Combining the two parts in (\ref{E6}), we find
\beq
    \int_0^\beta\! d^4y\,\ex{{\bf E}^2(x){\bf E}^4(y)} 
    = - {10\pi^4T^8\over 3^3\cdot 5^2}  
\eeq
in agreement with the first term of (\ref{t44}) in the main text. The other terms can then 
be derived along the same lines.

\section*{Appendix C}

From the quantization of the photon field between two parallel plates in Section 3, 
it is straightforward to construct the full propagators in the Coulomb gauge.
For the electric field we find for the $x$-component
\beq
    \ex{E_x(x)E_x(x')} &=& {2\over L}\int\!{d^2k_T\over (2\pi)^2}\int\!{dk_0\over 2\pi}
    \sum_{n=1}^\infty {i\over k_T^2}{(k_0k_y)^2 + (k_xk_z)^2\over K^2 + i\e} \nn \\
    &&\sin(k_zz)\sin(k_zz')e^{i\bk_T\cdot(\bx_T - \bx_T') - ik_0(t - t')}    \label{ExEx}
\eeq
where $K^2 = k_0^2 - k_T^2 - k_z^2$ with $k_z = n\pi/L$. The expression for the $y$-components 
has the same form with $k_x$ and $k_y$ interchanged in the numerator while the propagator for 
the $z$-component is
\beq
    \ex{E_z(x)E_z(x')} &=& {2\over L}\int\!{d^2k_T\over (2\pi)^2}\int\!{dk_0\over 2\pi}
    \sum_{n=1}^\infty {ik_T^2\over K^2 + i\e} \nn \\
    &&\cos(k_zz)\cos(k_zz')e^{i\bk_T\cdot(\bx_T - \bx_T') - ik_0(t - t')}     \label{EzEz}
\eeq
The magnetic propagators have the same general structure. There are also non-diagonal 
propagators between different components of the electric and magnetic fields.

When the two points in the propagators approach each other we get the field 
expectation values. The integral over $k_0$ can then be done by contour integration. For 
instance, for the $x$-component of the electric field we find from (\ref{ExEx})
\beq
    \ex{E_x^2(\bx)} = {1\over L}\int\!{d^2k_T\over (2\pi)^2}\sum_{n=1}^\infty
                      {k_y^2 + k_z^2\over \omega} \sin^2(k_zz)
\eeq
where now $\omega = \sqrt{k_T^2 + k_z^2}$. This is in agreement with the corresponding
expression (\ref{ETET}) in the main text.

Bordag {\it et al} \cite{Bordag} have succeeded in deriving a compact expression for all these
field propagators based on quantization in the Lorentz gauge $\del_\mu A^\mu = 0$. 
With the plates normal to the $z$-axis at the coordinate values $z = a_0$ and $z = a_1$, they
found
\be
         \langle A_{\mu}(x)A_{\nu}(y) \rangle = i \overline{D}_{\mu\nu}(x,y) 
        = \frac1{2}\int \frac{d^3\tilde{p}}{(2\pi)^3} 
          \frac{P_{\mu\nu}(p)}{\Gamma(p)}
          e^{i p_{\alpha}(x-y)^{\alpha}} e^{i \Gamma \mid x_3 - a_i \mid}
         \left( h^{-1} \right)_{ij} e^{i \Gamma \mid y_3 - a_j \mid} 
\ee
with $d^3\tilde{p} = dp_0 dp_1 dp_2$ and $p_{\alpha}x^{\alpha} = p_0 x_0 - p_1 x_1 - p_2 x_2$.
Here $P_{\mu\nu}$ is the projection operator
\be
        P_{\mu\nu}(p) = { \left\{  \begin{array}{c c}
                g_{\mu\nu} - \frac{p_{\mu} p_{\nu}}{\Gamma^2} &
                   \mbox{for $\mu$, $\nu$ $\neq$ $3$.} \\
                                                        &        \\     
                       0                             & \mbox{for $\mu=3$ or $\nu=3$.}
                                \end{array} \right. }
\ee
and $\Gamma(p) = \sqrt{p_0^2 - p_1^2 - p_2^2}$. The position of the plates enter also through the
matrix $h_{ij}$ with the inverse
\be
        \left( h^{-1} \right)_{ij} = \frac{i}{2 \sin\Gamma\mid a_0 - a_1 \mid}
                { \left(  \begin{array}{c c}
                                  e^{- i \Gamma \mid a_0 - a_1 \mid} &  -1 \\ 
                                        &       \\
                                  -1 & e^{- i \Gamma \mid a_0 - a_1 \mid}
                                \end{array} \right) }
\ee
In the following we will take $a_0 = 0$ and $a_1 = L$ as in the main text. 

The field expectation values can now be obtained by taking different derivatives of the propagator and
then let the field points coincide. In order to regularize the ensuing divergences, we take the
fields in the two points $(x^{\alpha}, z)$ and $(y^{\alpha}, z + \epsilon)$ where we let 
$x^\alpha \ra y^\alpha$ and leave the regulator $\epsilon$ infinitesimally small, but finite. For
instance, when we calculate the fluctuations in the $x$-component of the electric field we encounter
the expectation value
\be
        I_{-} &=& \langle \partial_0 A_1 \partial_0 A_1 \rangle \equiv
          \lim_{x^{\alpha} \rightarrow y^{\alpha}}
        \frac{\partial}{\partial x_0} \frac{\partial}{\partial y_0}  \langle 
         A_1(x) A_1(y)\rangle \\
        &=& \frac1{2} \int \frac{d^3 \tilde{p}}{(2\pi)^3} 
            \frac{p_0^2}{\Gamma} \left( - 1 - \frac{p_1^2}{\Gamma^2} \right)  
          e^{i \Gamma \mid z - a_i \mid} 
                \left( h^{-1} \right)_{ij} e^{i \Gamma \mid z + \epsilon - a_j \mid}  
\ee
By the Euclidean transformation $p_0 \ra i p_4$  under which $\Gamma \ra i\gamma$ this becomes
\be
        I_{-} &=& \frac1{2} \int \frac{dp_4 dp_1 dp_2}{(2\pi)^3} 
            \frac{p_4^2}{\gamma} \left( 1 - \frac{p_1^2}{\gamma^2} \right)  
         e^{- \gamma \mid z - a_i \mid} 
                \left( h^{-1} \right)_{ij} e^{- \gamma \mid z + \epsilon - a_j \mid} 
\ee
where now $0 < z < L$. Introducing spherical integration variables we then obtain
\be
        I_{\pm}&=&  \frac1{15\pi^2} \int_0^{\infty} d\gamma \hspace{1mm}
       \frac{\gamma^3}{2 \sinh(\gamma L)} \left[ 
                e^{-\gamma (2z+\epsilon-L)} \pm  e^{-\gamma (L-\epsilon)} \pm
        e^{-\gamma (L+\epsilon)} 
                + e^{-\gamma (L - 2z - \epsilon)} \right] 
\ee
after integrating over angles. While the function $I_{-}$ is needed for evaluation of 
$\langle E_x^2 \rangle$, $\langle E_y^2 \rangle$ and $\langle B_z^2 \rangle$, we have here also given
the related function $I_{+}$ which will be used to calculate $\langle E_z^2 \rangle$, 
$\langle B_x^2 \rangle$ and $\langle B_y^2 \rangle$. The remaining integrals are given in \cite{GR},
\be
        \int_0^{\infty} dx \hspace{1mm} \frac{x^{\mu-1} e^{-\beta x}}{\sinh(x)}
        = 2^{1-\mu} \Gamma\left(\mu\right) \zeta_H(\mu,(\beta+1)/2)
\ee
where $\zeta_H(s,x)$ is the Hurwitz zeta-function 
\be
       \zeta_H(s,x) = \sum_{n=0}^{\infty} \frac1{(n + x)^s}
\ee 
Thus, we have the results
\be
         I_{\pm} &=& \frac{1}{40\pi^2 L^4} \left[ \zeta_H(4,z/L) \pm \zeta_H(4,\epsilon/2L)  
         \pm \zeta_H(4,-\epsilon/2L)  +  \zeta_H(4,1 - z/L) \right] 
\ee
In the limit $\epsilon \rightarrow 0$, it follows from the above definition that
\be
         \zeta_H(s,\epsilon) = {1\over \epsilon^s} + \zeta(s) + {\cal O}(\epsilon)
\ee
when $s > 1$. We then have
\be
         I_{\pm} &=& \frac{1}{40\pi^2 L^4} \left[\zeta_H(4, z/L) + \zeta_H(4,1 - z/L)   
        \pm \frac{32L^4}{\epsilon^4}  \pm \frac{\pi^4}{45}  \right] 
\ee
For integer $s = k$ the remaining zeta-functions can now be expressed in terms of the digamma
function by the relation
\be
        \zeta_H(k,x) = \frac{(-1)^k}{(k-1)!} \frac{d^{k-1}}{dx^{k-1}} \psi(x)
\ee
In addition, using the identity  $\psi(x) - \psi(1-x) = - \pi \cot (\pi x)$, it follows that
\be
         \zeta_H(4,x)  +    \zeta_H(4, 1 - x)  = - \frac{1}{6}
           \frac{d^3}{dx^3} [\pi \cot (\pi x)] = 
        \frac{\pi^4}{3} \left( \frac{3}{\sin^4(\pi x)} - \frac{2}{\sin^2(\pi x)} \right)
\ee
With $\theta = \pi z/L$ as in the main text we can then write the result as 
\beq
        I_{\pm} = \pm \frac{4}{5 \pi^2 \epsilon^4} 
                + \frac{\pi^2}{120L^4}\left[\pm \frac1{15} + F(\theta)\right]
\eeq
where the function $F(\theta)$ is defined as in (\ref{F}).
This gives $\langle \partial_0 A_1 \partial_0 A_1 \rangle =  I_{-}$. Similarly we obtain
$\langle \partial_1 A_0 \partial_1 A_0 \rangle = 4 \langle \partial_0 A_1 \partial_1 A_0 \rangle 
= I_{-}$. Combining such terms, we then have 
\be
        \langle E_x^2 \rangle = \langle \partial_0 A_1 \partial_0 A_1 + 2
         \partial_0 A_1 \partial_1 A_0 +  \partial_1 A_0 \partial_1 A_0 \rangle = {5\over 2}I_- 
         =\frac{\pi^2}{48L^4} \left[-\frac1{15} + F(\theta)\right]
\ee
when we discard the divergent, distance-independent term. 
Both $\langle E_y^2 \rangle$ and $\langle B_z^2 \rangle$ can be evaluated in terms of $I_{-}$. For 
example, in the calculation of $\langle B_z^2 \rangle$ we need the expectation value
\be
         \langle \partial_1 A_2 \partial_1 A_2 \rangle &=& \lim_{x^{\alpha}\rightarrow y^{\alpha}}
        \frac{\partial}{\partial x_1} \frac{\partial}{\partial y_1}  \langle A_2(x) A_2(y)\rangle \\
        &=& - \frac1{2} \int \frac{dp_4 dp_1 dp_2}{(2\pi)^3} 
          \frac{p_1^2}{\gamma} \left( 1 - \frac{p_2^2}{\gamma^2} \right)e^{- \gamma \mid z - a_i \mid} 
                \left( h^{-1} \right)_{ij} e^{- \gamma \mid z + \epsilon - a_j \mid} =  -I_-
\ee
In this way we find that  $\langle E_x^2 \rangle = \langle E_y^2 \rangle = - \langle B_z^2 \rangle $
which agrees with what is obtained in the Coulomb gauge. Exactly along the same lines we also derive
$\langle E_z^2 \rangle = - \langle B_x^2 \rangle = - \langle B_y^2\rangle$ in terms of $I_+$, i.e. 
\be
\langle E_z^2 \rangle = \frac{5}{2} I_{+} = \frac{\pi^2}{48L^4} \left[\frac1{15} + F(\theta)\right]
\ee
when we again drop the divergent piece which represents the field fluctuation in empty space without
any disturbing plates.

\end{document}